# Benefits, Challenges and Contributors to Success for National eHealth Systems Implementation: A Scoping Review

James Scheibner,[1,2] Joanna Sleigh,[1] Marcello Ienca,[1] and Effy Vayena[1,3]

## Abstract:

**Objective:**

Our scoping review aims to assess what legal, ethical, and socio-technical factors contribute or inhibit the success of national eHealth system implementations. In addition, our review seeks to describe the characteristics and benefits of eHealth systems.

**Materials and Methods:**

We conducted a scoping review of literature published in English between January 2000 and 2020 using a keyword search on five databases; PubMed, Scopus, Web of Science, IEEEXplore, and ProQuest. After removal of duplicates, abstract screening and full-text filtering, 86 articles were included from 8276 search results.

**Results:**

We identified 17 stakeholder groups, 6 eHealth Systems areas, and 15 types of legal regimes and standards. In-depth textual analysis revealed challenges mainly in implementation, followed by ethico-legal and data related aspects. Key factors influencing success include promoting trust of the system, ensuring wider acceptance amongst users, reconciling the system with legal requirements and ensuring an adaptable technical platform.

**Discussion:**

Results revealed support for decentralised implementations because they carry less implementation and engagement challenges than centralised ones. Simultaneously, due to decentralised systems' interoperability issues, federated implementations (with a set of national standards) might be preferable.

**Conclusion:**

This study identifies the primary socio-technical, legal and ethical factors that challenge and contribute to the success of eHealth system implementations. This study also describes the complexities and characteristics of existing eHealth implementation programs, and surmises suggested guidance for resolving the identified challenges.

[1] Health Ethics and Policy Laboratory, Department of Health Sciences and Technology, ETH Zürich, Zürich, Switzerland

[2] College of Business, Government and Law, Flinders University, Adelaide, Australia

[3] Corresponding author. Email address: effy.vayena@hest.ethz.ch

## Background:

eHealth refers to a growing variety of platforms, from complex national programs to mobile health applications.[1,2] Just as we have seen the diversity of systems and technologies evolve in the last years, so too has the research sector increasingly framed eHealth systems as a means to promote accessibility, interoperability, security, and operationally efficiency. Increasingly, eHealth (defined as the use of information communication technologies for health[3]) is portrayed as a solution that can resolve many current problems with healthcare. These include the rising costs of treating aging populations and chronic diseases, inaccuracy of records and treatment due to human error, and fragmented delivery of healthcare. Accordingly, the problems with healthcare often get characterised as “wicked problems”, or problems with fragmented and contradictory requirements that are difficult to solve.[4]

While the use of information communication technology in healthcare undeniably has benefits, attempts to implement these programs has led to significant problems. Rather than solving the issues they sought to address, in a cyclical fashion, these eHealth implementations often become "wicked problems".[5] A prominent example is the United Kingdom's National Program for IT (NPfIT), designed to create a nationwide integrated electronic patient record system. The purported benefit of this program was to allow National Health Service (NHS) physicians to access patient records no matter which NHS practice or hospital the patient visited. However, NPfIT suffered severe delays due to a range of implementation, financial, and legal issues, and was eventually dismantled in September 2011.[6] The Dossier Médical Partagé in France encountered similar problems. However, the equivalent national electronic health record program in Estonia reports to have had higher rates of patient and physician uptake.[7,8] With diverging outcomes of large scale eHealth projects, an over and understanding of their benefits, challenges and contributors to success would help for future planning of such a project.

Acknowledging that much can be learned from recent eHealth initiatives, the objective of this scoping review was to explore the relevant literature describing eHealth systems implementations. A scoping review can help to identify the main areas of a research field that has not been comprehensively studied both at a broad and in-depth level.[9] Specifically, we seek to describe the characteristics of reported national health information systems and identify the socio-technical, legal and ethical factors that challenged or led to the success of such programs. For this purpose, we used references to national or regional health information systems platforms in a federal system as inclusion criteria. Several scoping reviews have already examined monitoring and governance of national electronic health records and eHealth strategies.[10–13] However, our review attempts to identify the specific factors that are reported as benefits, challenges and factors in the implementation of these systems. Results from this study will help future implementation projects identify what might be the critical factors in the success and failure of these projects.

## Methods:

Our scoping review followed the PRISMA Extension for Scoping Reviews (PRISMA-ScR).[14] As such, we report the method used according to the following steps: (1) identifying the research questions, (2) identifying the relevant literature, (3) selecting articles, (4) analysis of articles, (5) collating, summarising and reporting results.

### Research Questions

With this scoping review, we sought to answer two research questions:

1. What are the main characteristics of eHealth systems?
2. What are the benefits, challenges and contributors to success for eHealth systems implementation on a national and regional level?

### Identifying Relevant Literature

We developed the search strategy through discussions with the broader research team for the Data Protection and Personalised Health (DPPH) project, which included two senior academics in bioethics and a senior researcher in law. Table 1 displays the search strategy and Figure 1 provides an overview of the six stages of screening.

We performed a search across five databases using various key terms relating to the review's central concepts of 'eHealth', 'intervention' and 'location' (Table 1.). There was no restriction on the type of publication, study design, or methods used. The search resulted in 8276 articles, which after duplicates were manually removed left 5728. Next, we screened the full text of the resulting papers based on the search strategy presented in Table 1 and Figure 1. Papers were included if they had the relevant search terms in their full text, were published between January 2000 and 2020, were written in English, and were either a book chapter or a journal article. We excluded letters to the editor, correspondences, comments, and editorials. Papers were excluded if they did include the "eHealth" and "Intervention" key terms or describe the implementation in their abstract or title. Retracted papers were also excluded.

Further, for inclusion papers had to either discuss and assess, retrospectively or prospectively, health information system implementations. Selecting this spectrum of different studies helped us adopt a narrative approach to assessing the success or failure of different programs.[15] The challenges or factors of success described were not limited to either quantitative or qualitative measures. For example, levels of participation in a national health record system or a cost analysis showing an increase in healthcare efficiency could be considered quantitative measures of success. By contrast, availability, functionality or perceptions about the trustworthiness of a national health record system could be considered a qualitative measure of success.[16]

**Table 1. Search Strategy**

| Process | Detail |
|---|---|
| Databases | PubMed, Scopus, Web of Science, IEEEXplore, and ProQuest |
| Search Terms | eHealth terms: "Electronic health system" or "Electronic medical system" or "Health information system" "Electronic health record" or "Electronic medical record" or "health data" or "biomedical data" or "patient data" or "biomedical research"<br>AND<br>Intervention terms: "Centralised" or "Centralized" or "Decentralised" or "Decentralized" or "Federated" or "Blockchain" or "Distributed" or "Governance" or "E-Portal" or "Deployment"<br>AND<br>Location terms: "National" or "Nationwide" or "Regional" or "Jurisdictional" or "Consortium" or "Government" |
| Type of Study | Journal Article or Book Chapter |
| Time range | January 2000-January 2020 |
| Limits | English |
| Inclusion | Abstract describes implementing an eHealth system and challenges or success factors. |
| Exclusion | Articles that were retracted |

**Article Selection**

After the initial screening, a total of 622 remained for full-text review (Figure 1). For these articles identified, we used the Zotero reference management software to extract the year of publication, authors, title and abstract, full-body text (if available) and references. We then screened the full texts for the following inclusion and exclusion criteria.

Included were full-text articles in English that focused on national health information system platforms and implementation strategies. Formats included were implementation studies, empirical or comparative studies, or opinion, technical, survey-based, doctrinal and legal articles. Further, articles needed at least one paragraph on two of the following factors: ethical, legal or socio technical. Articles also had to have at least one paragraph on potential quantitative and qualitative measures of success and failure, including ex-ante observations about the successes and failures/challenges of a health information system, or ex-ante comparative observations about the success and failure of a health system.

We excluded articles that did not have full texts in English or were letters to the editor, systematic reviews and study protocols. We also excluded articles that only tangentially mentioned ethical, legal or socio-technical factors (for example, mentioning ethical or legal requirements as part of an ethical review), or that focused on other factors such as only on economic or technical ones. Also excluded were articles that did not have at least one paragraph on potential quantitative and qualitative measures of success and failure. Finally, we excluded articles that focused only on mobile health applications or medical devices. Our justification for this was that these devices are often implemented on an institutional basis, as opposed to software implementations that occur on a national or regional basis. Our overall approach was described in Figure 1.

**Figure 1.** PRISMA flow-chart displaying the search protocol for articles

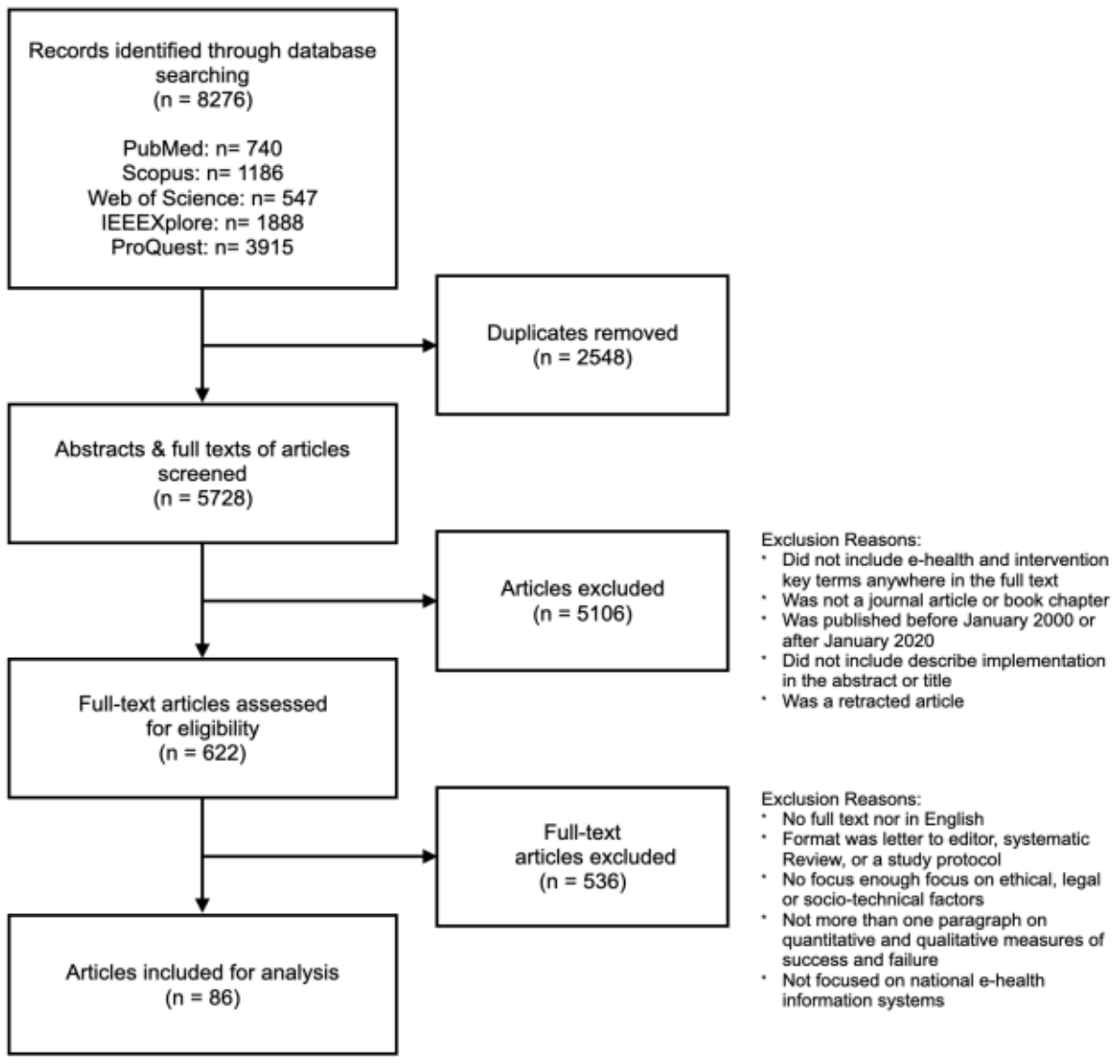

### Analysis of Articles

The next stage of the project was to code each of the articles quantitatively. In this process, we identified the characteristics of the eHealth systems described and the factors leading to success or failure. We inductively identified these categories using the thematic model suggested by Braun and Clarke.[17] Also, we conducted a thematic content analysis on specific categories identified from each of the articles. Table 2 provides details.

Coding was carried out in unison by two of the authors using the software MaxQDA. The authors coded phrases by sentence mentions (as opposed to full sentences). Except for legal regimes and regulations, the authors excluded footnotes, keywords, references and titles from the scope of coding. To capture legislation, we included footnotes and references in our analysis. The same authors as above coded the first ten documents separately to minimize subjective bias, then checked for congruence. Once these codes were reconciled, the authors split the remaining 77 documents and coded them independently. The authors then merged these documents together once coding was completed.

**Table 2. Inductive Coding Themes**

| Category | Description |
|---|---|
| Article type | Refers to the type of study performed. After initial coding, an *ex-ante* approach conducted in concert between the two coders revealed the following categories and subcategories of study type:<br>● Empirical Studies: analyse the effects of eHealth system implementation, such as stakeholder assessments, case studies; media analyses;<br>● Implementation studies: explain implementation of the systems, such as reviews; |
| | • Doctrinal studies: ethico-legal articles such as single jurisdiction analysis and comparative analysis papers. |
| Taxonomy of system | This category was inductively coded, and referred to the health information system as it was described within the document. For example, a distinction was drawn between electronic health systems (as a reference to a longitudinal record associated with a single individual) and electronic medical records (as a reference to a software system for storing electronic health records[18]). |
| Structure of System | As alluded to with respect to the level of health information systems, the authors identified three ways in which a health information system could be implemented. The first is a centralised or 'top down' model. Following this approach, the health information system is implemented by a government and rolled out to all hospitals and healthcare institutes. An example of this approach is the National Health Service Information for Health Strategy, a comprehensive rollout that was attempted in the United Kingdom following the electoral victory of Labour in 1997. This centralised approach has the advantage of increasing standardisation and procurement processes, increasing the likelihood of health records being used for secondary purposes. However, the National Health IT System implementation process has also been plagued with 'setbacks, misgivings, clinical unrest, delays, cost overruns and pairing back of promised functionality'.[19] Further, placing all health records in one location creates a single point of failure, increasing the impact of a potential data breach. |
| Stakeholders | The coders agreed that different stakeholder categories would be coded inductively from each article. These included all actors that might be involved in the creation of health information systems. Once we had coded these categories, we then grouped them into subcategories. |
| Country/ Jurisdictions | The coders agreed to inductively code both the jurisdiction of the authors and the jurisdiction of study. |
| Legal Regimes / Regulations | This category referred to any international guidelines and national or supranational legislation mentioned within the article. These values were separated in commas. |
| Benefits, Challenges & Contributors to Success | An inductive coding mechanism was used to identify any factors which could support or undermine implementation. As discussed previously, these three code categories were subject to thematic content analysis.[20] |

# Results

Our search and selection resulted in 86 articles for full-text analysis. Of these, 53 were empirical studies, such as qualitative or quantitative stakeholder assessments, case studies, media analyses or implementation reports. 16 articles were implementation studies describing a particular architecture or implementation strategy and the remaining 17 were ethico-legal articles.

## Description of eHealth systems

### *Taxonomy & structure of eHealth systems*

The articles described a diversity of eHealth systems, surmised by six categories: (1) *health data*, such as patient data, (2) *data collection*, like medical records or registries (3) *technologies*, meaning information communication technologies (4) *systems*, such as pharmacy information systems, (5) *healthcare services*, like e-prescription and telehealth, and (6) *networks*, such as clinical data research networks. Of these six areas, the majority of documents (n=53/86, 61%) made references to types of *data collections*. This category also had the most subcategories (Supplementary materials). The least reported category was *networks*, which had just four different subcategories and 49 instances across five documents. The supplementary details each category's subcategory and the sum of their occurrences across documents. Centralised eHealth systems appeared most frequently (n=38/86, 44%), followed by decentralised (n=24/86, s=58). Least reported was federated eHealth systems (n=12/86, s=21).

### *Stakeholders*

We identified 17 different groups of e-Health system stakeholders. The majority of articles (n=85/86) referenced the stakeholder groups of patients & the lay public as, along with primary health care providers (n=80/86), and healthcare organisations (n=78/86). Government and academic stakeholders were also often mentioned (both n=71/86), followed by information communication stakeholders (n=37/86), legal (n=37/86) and pharmaceutical stakeholders (n=33/86). Less frequently mentioned across all documents were data protection stakeholders and insurance organisations (both n=27/86).

**Figure 2. Distribution of Stakeholders**

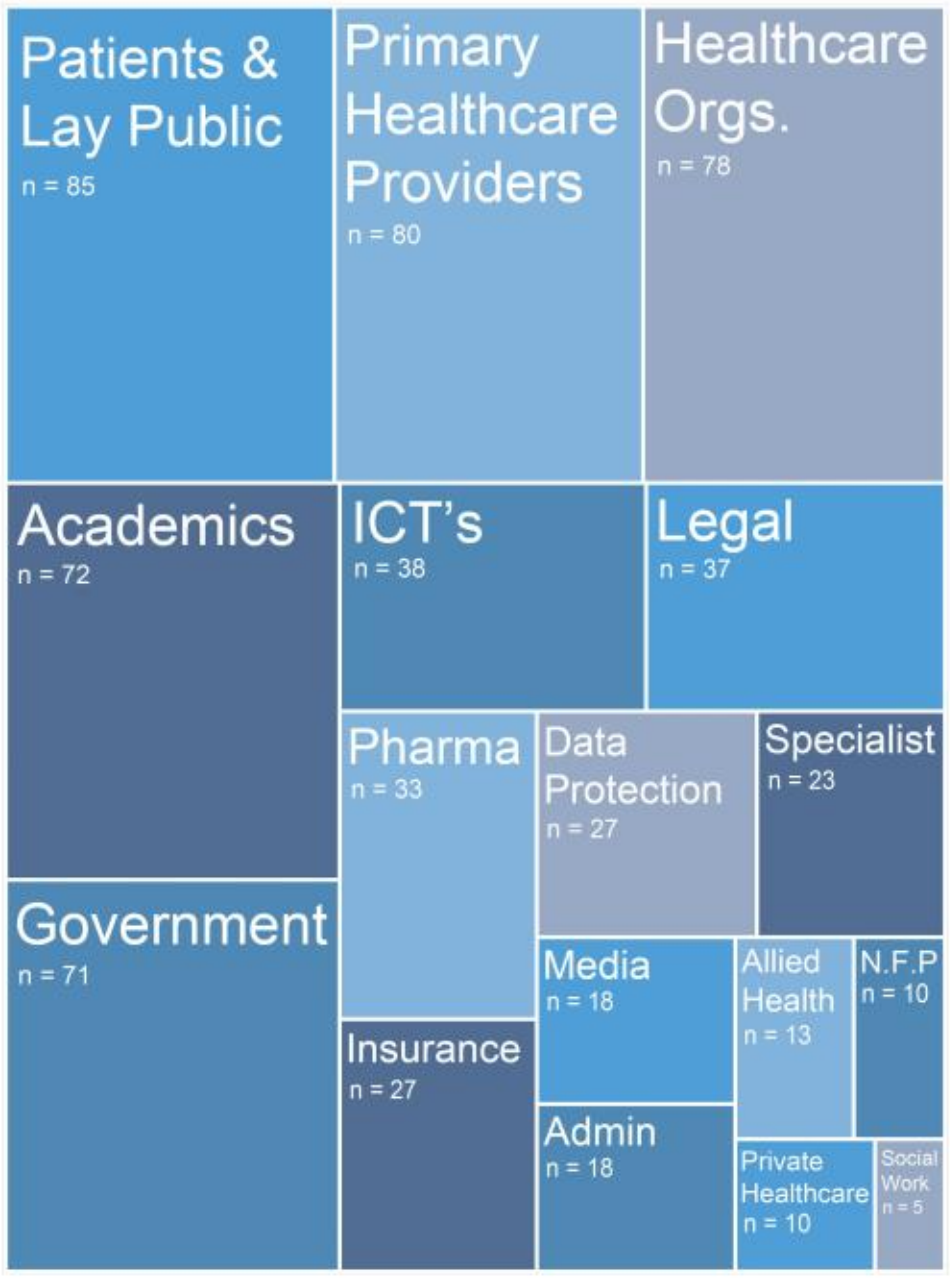

### *Countries & Jurisdictions*

Analyses of the documents revealed 48 different countries and jurisdictions involved in the development or implementation of eHealth systems. As shown in Figure 3, the three most mentioned countries were the UK, USA and Australia. Canada, the European Union, Denmark, Sweden and Singapore were also frequently mentioned.

**Figure 3. Jurisdictions included.**

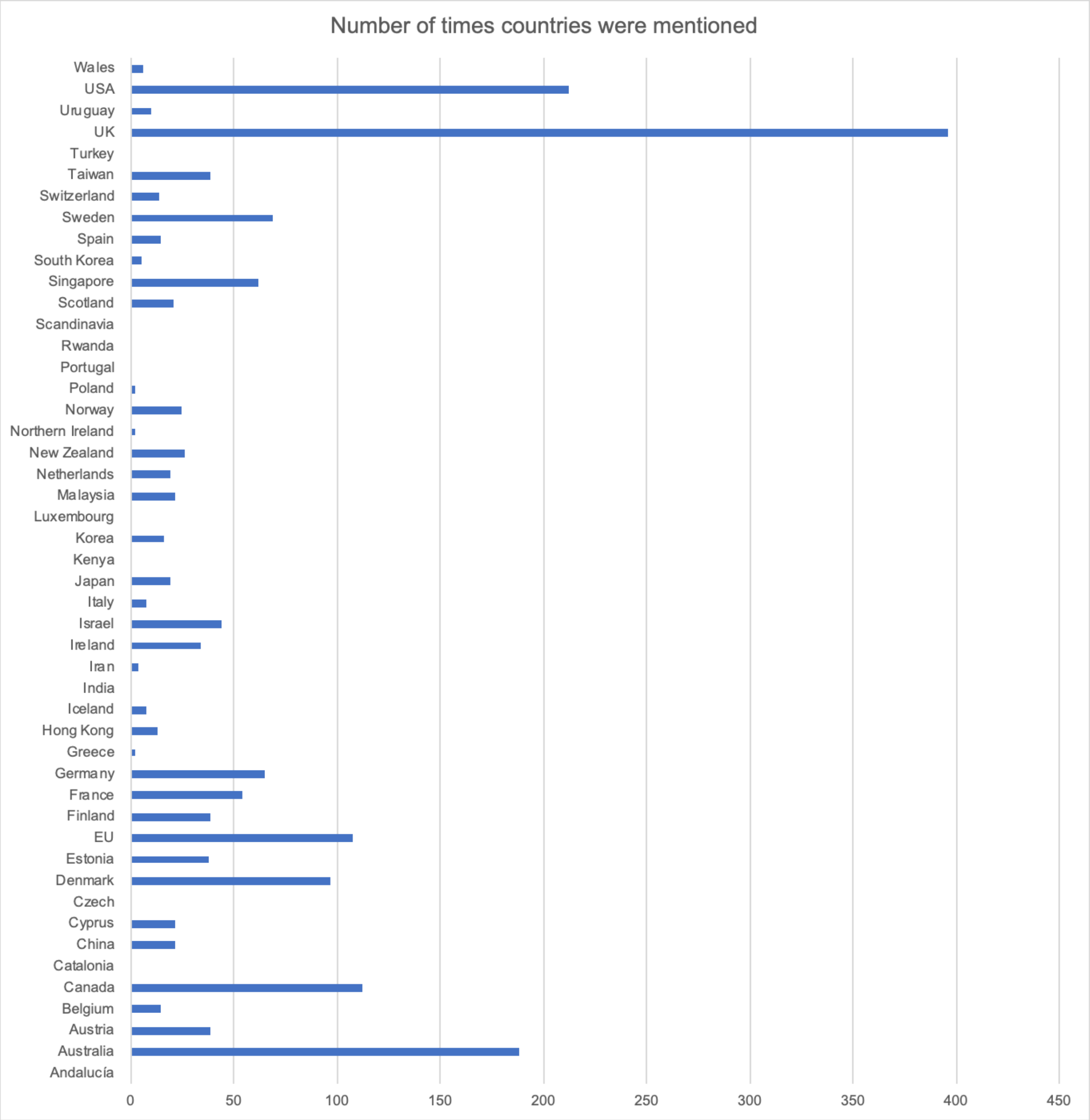

### *Legal Regimes & Regulations*

We coded and clustered legal regimes into 17 categories (supplementary materials). The most frequently reported were national data and privacy protection laws and legislations (n=29/86, s=96). Following were national electronic health record and health funding legislations (n=45/86, s=17), these provide rules on processing health records and often supersede data protection laws. Likewise, health funding laws are crucial for guaranteeing the ongoing sustainability and support of eHealth systems. Although there was some mention of supranational data protection frameworks, these were often mentioned in the context of national data protection laws.[21–33]

National health service laws (n=10/86, s=20) also offered guidelines on using and implementing eHealth systems.[34,35]

## Benefits, Challenges & Contributors to Success

### *Benefits of eHealth Systems*

The articles analysed reported a plethora of benefits to eHealth systems. Most frequently, they were said to improve the efficiency and effectiveness of healthcare coordination,[36,37] processes and delivery (n=57, s=156). For example, by "simplify[ing] administrative tasks",[38] helping with "optimising physical resources",[39] and "reducing duplication of services, realising operational efficiencies".[40] Many documents also described eHealth systems as a means to improve access and exchange of information and data (n=57/86, s=144).[41–46] Over half of the documents also reported that eHealth would improve the quality of care (n=48/86, s=107)[47], and support research and policy (n=43/86, s=92). Other perceived benefits included patient empowerment & engagement,[46,48–51] improvements to patient safety and data security[53], reduction of costs, better service monitoring[54] and generally increased ability to address challenges that emerge.[55] Table 7 provides details.

### *Challenges of eHealth Systems*

More prevalent than the benefits, however, were challenges. Overall, we coded challenges within five categories; implementation, legal and ethical, data-related, engagement, and software-related challenges (Figure 4).

Implementation challenges frequented most, and included conflicting stakeholder requirements (n=26/86, s=48),[56,57] difficulty demonstrating benefits (n=32, s=76/86),[58] financial issues (n=29/86, s=104), government, policy and political issues (n=18/86, s=28) as well as broader implementation challenges (n=45/86, s=238).[59]

Legal and ethical challenges also abounded. Most frequently reported were concerns about privacy (n=54/86, s=192),[60] and research ethics (n=15/86, s=86). The latter referred to challenges with ethics committee approvals,[61–64] ensuring representative samples, reporting results, protecting the rights of patients,[38,61,65,66] and consent /re-consent processes[67–70]. The articles we examined also identified patient autonomy, or the right of patients to make decisions about their medical care and overall health, as a significant challenge. In this context, patient autonomy specifically referred to patients controlling access to their electronic records, and whether an opt-in or opt-out consent was preferable.[59] Also, our sampled documents mentioned three categories of medical liability (n=17/86, s=42).[70] These were the liability of physicians and healthcare organisations to provide adequate medical care, device or product liability for eHealth systems, and patient responsibility for their health records.[21,71–73]

Over two-thirds of the documents reported data challenges (n=68/86, s=318). First, this related to challenges of ensuring data availability (n=50/86, s=148), meaning the ability or willingness to share data with other hospitals, healthcare institutions or research institutions.[61,74–76] For example, on a jurisdictional level was the challenge of data transfer across national or international borders.[62,68,77] Likewise, challenges arose regarding the availability of different data forms, such as metadata and data linkage.[37,62,77,78] The second data challenge was information quality (n=31/86, s=53) referring to poor quality, missing, or lost records and information.[19,79,80] The third data challenge was interoperability (n=43/86, s=117), in that using heterogeneous standards would limit data sharing.[42,44,81–83] Likewise, the use of "silo" or free text systems was said to undermine the free exchange of data.[58]

**Figure 4.** Sunburst diagram of eHealth system challenges and sub-categories according to no. of segments.

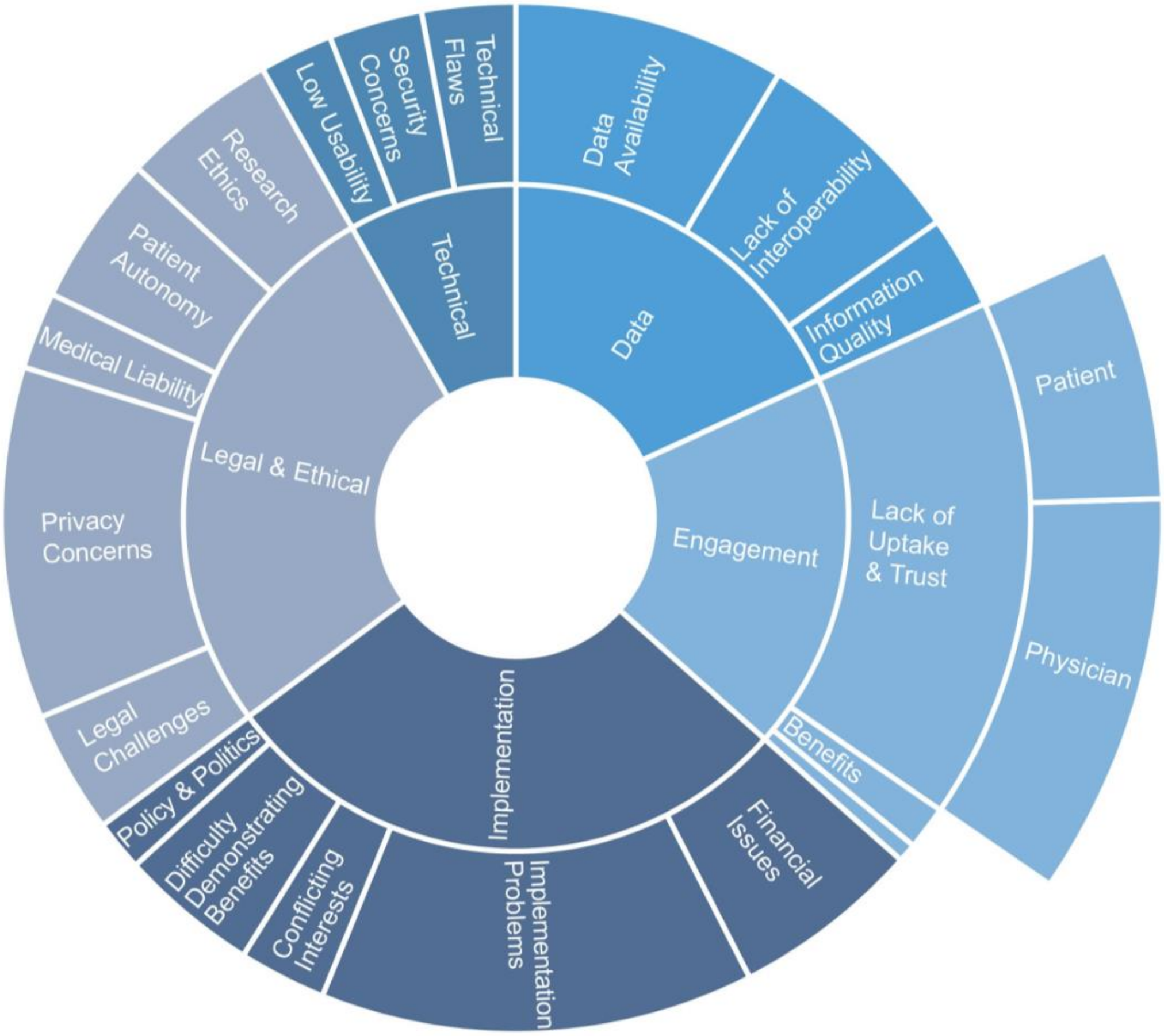

Stakeholder engagement was another major challenge. First, was the issue of lack of uptake. This involved a lack of patient uptake (n=33/86, s=108), which would undermine the use of patient data for secondary research purposes.[52,84,85] As well, physicians would refuse to use the systems because they did not trust or had had poor experiences,[86–89] with software frequently being attributed as a cause in physician burnout (n=41/86, s=176). In turn, this would limit the benefits that might flow from these systems.[90–93] Also impacting engagement was lack of access to benefits from using eHealth systems (n=12/86, s=23), such as from geographically isolated areas and developing countries,[40,94] and communication issues (n=8/86, s=10).

Finally, technical challenges associated with the systems themselves were mentioned (n=42/86, s=142). Out of these, security challenges were most frequent (n=24/86, s=51), and involved difficulties with security measures, including role-based access control and information security.[55,95] Further, these challenges were directly related to broader ethico-legal concerns about patient autonomy and privacy, as well as impacting the engagement of physicians and patients

### *Contributors to success*

We identified six categories of factors contributing to successful eHealth systems and their implementation. As shown in Table 9, the categories were socio-cultural, system-based, legal, implementation, and resources.

Most frequently highlighted as an underlying success factor was stakeholder engagement (n=58/86, s=218). Users' trust and support were integral. One article describes "*The adoption and diffusion of a large-scale IT programme necessitates the cultural and normative acceptance of a wide array of constituents, not merely within an organisation, but at the levels of the organisational field and wider society*".[96,97] Many articles proposed methods of incentivisation alongside skills training and the provision of guidance to foster support, trust and shared ownership of the system. Multi-disciplinary collaborations too were advised, for example by engaging with patient representatives, having public involvement at every stage, or by letting government policymakers set the agenda. Communicating about lessons learned and sharing past experiences was also suggested to help speed up implementation and reduce error.

The majority of articles also emphasised several system qualities (n=64/86, s=212). eHealth systems needed to be flexible, intuitive, accessible, transparent, reliable, interoperable, and safe. Having context-sensitive security processes and robust data governance structures would "*demonstrate trustworthiness and sustain the social license*".[97–99] Data accuracy and consent mechanisms were needed that served user needs by promoting informational autonomy.[83,100,101] Recommended were also decentralised systems because "*centralisation causes paralysis*"[102].

Regarding legal contributors to success, over half of the documents reported the importance of abiding by and supporting the development of standards, protocols, guides and recommendations (n=45/86, s=134).[57] Also important were good implementation practices, such as defining goals, having a realistic timeline, and iterative evaluation processes, (n=40/86, s=149).[56] Interestingly, the least reported factor of success was resources (n=32/86, s=82).

## Discussion

This scoping review paints the first comprehensive picture of eHealth system implementations, as documented in the current literature. To begin with, the results demonstrate that eHealth systems are characterised by complexity and diversity, in their connection of divergent data types, data collections, different technologies, systems and networks, all made up of different actors. Beyond just referring to electronic health records, eHealth systems include patient-controlled records and eHealth cards, designed to make healthcare more accessible for patients. At the same time, eHealth refers to computerised physician order entry, decision support, and laboratory information systems, all of which help physicians with their work. Indeed, as the results show, eHealth systems promise benefits to all stakeholders, particularly medical practitioners, patients, but also policymakers, researchers and the general public.

Interestingly, through this scoping review we see a decline in support for centralised eHealth implementations, and instead a growth in support for decentralised and federated platforms. As mentioned previously, the earliest implementations of national electronic health record systems were those implemented in the United Kingdom. However, these early implementations were plagued by numerous project management issues, and security concerns about storing data in one location. These shortcomings led to distrust and a lack of uptake from both physicians and patients.[19,103] For this reason, the United Kingdom abandoned their centralised approach and have shifted a decentralised one. Notably, decentralised systems have a higher degree of patient engagement.[81] However, these systems also face challenges in terms of integration and interoperability. Therefore, federated systems may offer a useful compromise in guaranteeing the security of patient data whilst ensuring that healthcare systems remain interoperable.

Regarding challenges, our analysis demonstrated that technical shortcomings such as system architecture are not necessarily the greatest impediment to success. Nonetheless, technical issues have a flow on effect, such as leading to low usability and security concerns. One means to avoid these two issues, as recommended by the articles analysed, is to ensure from the onset that the eHealth system implemented is intuitive, interoperable, transparent, accessible, safe and reliable. Ensuring these system qualities, along with robust data governance structures and context-sensitive security contributes to success.

Of greatest challenge to eHealth systems are ethico-legal factors, particularly privacy and research ethics concerns, such as informed and broad consent, secondary uses of data and return of results. To address these, implementation projects need to actively foster support, trust and ownership of the system amongst all actors, while also incentivising engagement and providing training and guidance to actors involved. The articles analysed also stressed the importance of ensuring eHealth systems comply with the laws of the jurisdiction in which they operate. Further, some jurisdictions had implemented national electronic health records legislation that determined the use of these systems (such as Australia and Germany). Ensuring that all systems are compliant with this legislation, which may supervene national privacy laws, and to ensure that there is no mismatch between technical and legislative solutions, however, can also be challenging. Finally, legislation on healthcare system funding is important to ensure the ongoing financial sustainability of eHealth systems.

Outside of implementation and legal challenges, the high rate at which data challenges were mentioned indicates the necessity for interoperable standards and vocabularies. In turn, this need may make federated systems that supply a set of national standards preferable to decentralised systems. Coiera (2011) notes that a "middle out" national strategy might be preferable for certain health data types, such as summary care records, which are more "liquid" than other records such as prescription information.[19] Likewise, the need for data availability and interoperability may inform the technical design for eHealth systems. For example, Doods et al (2014) and Hailemichael et al (2015) describe a distributed learning platform for conducting privacy preserving computations on patient data. Both of these platforms crucially allow for these computations to be performed without the data leaving the facility where they are stored.[75,104] These technical solutions may have an impact on another major challenge identified in this study, namely the lack of patient or physician uptake.

**Strengths & limitations**

One potential limitation of this study is that our search terms may have failed to capture all studies that would have met the inclusion criteria. Another limitation is that we focused only on articles published in the last twenty years. While this ensures our review reflects current practice, given the rapidly changing nature of technology and cultural norms, some of the issues captured form two decades ago may no longer be relevant. Further, there may have been grey literature or articles in languages other than English relevant to our review, however their inclusion was beyond the scope of this study. Despite these limitations, given the methodological approach taken and the inclusive search strategy, we believe this scoping review provides an overview of the characteristics, benefits, challenges and factors contributing to the success of eHealth systems internationally. Finally, one of the difficulties in measuring the successes and challenges of eHealth implementations is that these projects are often multiyear initiatives. Our scoping review did not include longitudinal changes in benefits, challenges and success factors. Therefore, ongoing monitoring of eHealth implementations is crucial for measuring any benefits and challenges[13].

# Conclusion

This review provides an overview of the benefits, challenges and factors to success for respective implementations. The analysis of results brings us to the conclusion that eHealth systems provide a diversity of benefits. In particular, eHealth systems can contribute to improving access and exchange of information and data. Further, they can improve the quality of care, reduce costs, support research and policy, safeguard patient empowerment and safety. However, for these benefits to actualize, it is critical to focus on their implementation which requires attention to more than just the technologies themselves.

## Funding and Competing Interests:

This work was partially funded by the Personalized Health and Related Technologies Program (grant 2017-201; project: Data Protection and Personalized Health) supported by the Council of the Swiss Federal Institutes of Technology. The authors declare that there are no competing interests.

## Author Contribution:

JScheibner, MI, JSleigh and EV contributed to the conception of the study. JScheibner, MI and tested the protocol. JScheibner and JSleigh performed the analysis and wrote the first draft of the manuscript. JSleigh designed the tables and diagrams. All authors contributed to the manuscript revision, read, and approved the submitted version.

# Supplementary materials:

**Search Strategy:**

1. Logic Grid, Search Strings and Initial Retrieval Results:

| **System** | **Intervention** | **Location** |
|---|---|---|
| AND | AND | AND |
| “Electronic medical system” | Centralised | National* |
| “Electronic healthcare system” | centralized | Nationwide* |
| “Electronic health record*” | decentralized | Regional* |
| “Electronic medical record*” | decentralised | Jurisdiction* |
| “health data*” | federated | Consortium* |
| “patient record*” | distributed | Government* |
| “patient data*” | Blockchain | |
| “Health information system*” | Governance | |
| “biomedical data*” | e-Portal* | |
| “biomedical research*” | deployment | |

a). PubMed:

When converted into PubMed, this provides the following search string:

**((((((((((((electronic medical system[Title/Abstract]) OR electronic healthcare system[Title/Abstract]) OR electronic health record*[Title/Abstract]) OR electronic medical record*[Title/Abstract]) OR health data*[Title/Abstract]) OR patient record*[Title/Abstract]) OR patient data*[Title/Abstract]) OR health information system*[Title/Abstract]) OR biomedical data*[Title/Abstract]) OR biomedical research*[Title/Abstract])) AND ((((((((((centralised[Title/Abstract]) OR centralized[Title/Abstract]) OR decentralised[Title/Abstract]) OR decentralized[Title/Abstract]) OR federated[Title/Abstract]) OR distributed[Title/Abstract]) OR blockchain[Title/Abstract]) OR governance[Title/Abstract]) OR e-portal*[Title/Abstract]) OR deployment[Title/Abstract])) AND ((((((national*[Title/Abstract]) OR nationwide*[Title/Abstract]) OR regional*[Title/Abstract]) OR jurisdiction*[Title/Abstract]) OR consortium[Title/Abstract]) OR government[Title/Abstract])**

Using this search string in May 2020, **740** results were returned.

b). Scopus

(TITLE-ABS(*"electronic medical system"*) OR TITLE-ABS(*"electronic healthcare system"*) OR TITLE-ABS(*"electronic health record*"*) OR TITLE-ABS(*"electronic medical record*"*) OR TITLE-ABS(*"health data*"*) OR TITLE-ABS(*"patient record*"*) OR TITLE-ABS(*"patient data*"*) OR TITLE-ABS(*"health information system*"*) OR TITLE-ABS(*"biomedical data*"*) OR TITLE-ABS(*"biomedical research*"*)) AND (TITLE-ABS(*centralised*) OR TITLE-ABS(*centralized*) OR TITLE-ABS(*decentralised*) OR TITLE-ABS(*decentralized*) OR TITLE-ABS(*federated*) OR TITLE-ABS(*distributed*) OR TITLE-ABS(*blockchain*) OR TITLE-ABS(*governance*) OR TITLE-ABS(*"e-Portal*"*) OR TITLE-ABS(*deployment*)) AND (TITLE-ABS(*national**) OR TITLE-ABS(*nationwide**) OR TITLE-ABS(*regional**) OR TITLE-ABS(*jurisdiction**) OR TITLE-ABS(*consortium**) OR TITLE-ABS(*government**))
A search conducted within this string in May 2020 returned 1186 results.

c) Web of Science

Use exact quotes for this search.

(((TS=("electronic medical system" OR "electronic healthcare system" OR "electronic health record"* OR "electronic medical record"* OR "health data"* OR "patient record"* OR "patient data"* OR "health information system"* OR "biomedical data"* OR "biomedical research"*)) AND (TS=(centralised OR centralized OR decentralised OR decentralized OR federated OR distributed OR blockchain OR governance OR e-Portal*)) AND (TS=(national* OR nationwide* OR regional* OR jurisdiction* OR consortium* OR government*))))

A search conducted with May 2020 returned 547 articles.

d) IEEE Xplore

Note that this search string required treating the location and intervention as optional to one another.

**((((((((((("Abstract":"electronic medical system") OR "Abstract":"electronic healthcare system") OR "Abstract":"electronic health record") OR "Abstract":"electronic medical record") OR "Abstract":"health data") OR "Abstract":"healthcare data") OR "Abstract":"health information systems") OR "Abstract":"patient record") OR "Abstract":"patient data") OR "Abstract":"biomedical data") OR "biomedical research")**

AND

**( ((((((((((("Abstract":centralised) OR "Abstract":centralized) OR "Abstract":decentralised) OR "Abstract":decentralized) OR "Abstract":federat*) OR "Abstract":distribut*) OR "Abstract":blockchain) OR "Abstract":governance) OR "Abstract":e-portal) OR "Abstract":deploy*) OR "Abstract":implement*)**

OR

**((((("Abstract":nation*) OR "Abstract":regional) OR "Abstract":jurisdiction) OR "Abstract":consortium) OR "Abstract":government) )**

Running this search string in May 2020 returned 1910 articles. Only 1888 of these could be stored in Zotero for screening due to indexing errors when attempting to download the articles into Zotero.

e) ProQuest:

(ab("electronic medical system" OR "electronic healthcare system" OR "electronic health record" OR "electronic medical record*" OR "health data" OR "patient record" OR "patient data" OR "healthcare data" OR "health information system" OR "biomedical data" OR "biomedical research") OR ti("electronic medical system" OR "electronic healthcare system" OR "electronic health record" OR "electronic medical record*" OR "health data" OR "patient record" OR "patient data" OR "healthcare data" OR "health information system" OR "biomedical data" OR "biomedical research")) AND (ab(centralised OR centralized OR decentralised OR decentralized OR federat* OR distribut* OR blockchain OR governance OR e-portal OR deploy* OR implement*) OR ti(centralised OR centralized OR decentralised OR decentralized OR federat* OR distribut* OR blockchain OR governance OR e-portal OR deploy* OR implement*)) AND (ab(nation* OR regional OR jurisdiction OR consortium OR government) OR ti(nation* OR regional OR jurisdiction OR consortium OR government))

**Table 1. Taxonomy and Structure of eHealth systems**

| Category | Description | Docs (n) | Segments (s) |
|---|---|---|---|
| Data | Patient data; health data; hospital data | 53 | 152 |
| Data Collections | Clinical Registry; Computer-Stored Ambulatory Record; Continuity of Care Record ; Electronic Health Card; Electronic Health Record; Electronic Medical Record; Electronic Patient Records ; Electronic Personal health record; Electronic summary care record; European Health Insurance Card; Health Insurance Card; Health Professional Cards; Lifetime Health Record; Medical Records; National Clinical Registry; National Electronic Health Record; National Health Dataset; National Health Insurance Smart Cards; Nationwide Electronic Patient Database; Patient Accessible Electronic Health Record; Patient Medical Record; Personal Health Record; Personally Controlled Electronic Health Record | 79 | 3969 |
| Technologies | Health Information Technologies; Information Communication Technologies; Interoperable Enterprise Health Records Solution | 34 | 202 |
| Systems | Clinical Information Systems; Computerized Patient Record System; Computerized Physician Order Entry Systems; Decision Support Systems<br>Health Information Systems; Hospital Information System; Laboratory Information Systems; Nationwide Integrated Healthcare System; Patient Accessible Medical Record System; Pharmacy Information System; Radiology Information Systems; Systems for Health Care Transaction | 22 | 209 |
| Services | Computerised Practitioner Order Entry; eHealth; E-Prescription; Health Telematics; Online Health Record Service; Teleconsultation; Telehealth; Telemedicine | 43 | 540 |
| Networks | Clinical Data Research Networks; Health Information Network; National Health Data network; Patient-powered Research Networks | 5 | 49 |
| **System Structure** | | | |
| Centralised | Data or records are stored in a single centralised repository. On a national level, the program is implemented through mandating standards and software at a national level. | 38 | 137 |
| De-centralised | Data or records are stored at the health care centre or research institute where they were collected. Data sharing occurs when records are transferred between data custodians. Although health care centres or research institutes develop their own standards, | 24 | 58 |
| Federated | Encapsulates a hybrid between centralised and decentralised systems. This category includes the eHealth systems that are administered on a federal basis, where each state or region of the country has its own standards (such as in Australia, Denmark, Spain, or Switzerland). It can also encapsulate systems where individual hospitals share common standards and engage in joint analysis and processing, but keep their data separate. | 12 | 21 |

## Table 2. Types of legislation identified

| Categories | Examples | Docs (n) | Segments (s) |
|---|---|---|---|
| Data Protection and Privacy Laws (National) | Act 18331 of August 11 2008 (Uruguay), Data Act 1973 (Sweden), Datenschutzgesetz 2000 (Austria), Data Protection Act 1998 (UK), Data Protection Act 2000 (Netherlands), Health Insurance Portability and Accountability Act (United States), Law of 8 December 1992 on protection of private life in relation to personal data processing (Belgium), Patient Data Act 2008 (Sweden), Personal Information Act 2018 (Norway), Personal Information Protection and Electronic Documents Act (Canada), Privacy Act 1988 (Australia), Personal Data Protection Act 2010 (Malaysia), Personal Data Act 532/1999 (Finland) | 29 | 96 |
| National Electronic Health Record Law | Act 18211 of December 13 2007 (Uruguay), Act on Electronic Health Information Exchange 2013 (Netherlands), Act on Electronic Prescriptions 61/2007 (Finland), Act on the National Personal Data Registries for Health Care 556/1989 (Finland), Decree Law 98/2013 (Italy), eHealth Act 2015 (Germany), Digital Agenda 2011-2015 (Uruguay), ELGA Act 2012 (Austria), Health Information Systems Development Plan 2004 (Estonia), Legislative Decree no. 179/2012 (Italy), Legislative Decree no. 82/2005 (Italy), Legislative Decree no. 69/2013 (Italy), National Regulatory Framework (Sweden), Law of March 4, 2002 (Belgium) | 17 | 45 |
| Health Funding Law | Act on Private Health Care 152/1990 (Finland), Affordable Care Act 2010 (United States), American Taxpayer Relief Act 2012 (United States), American Recovery and Reinvestment Act 2009 (United States), General Health Care Act 1986 (Spain), Gesundheitsreformgesetz 2005 (Austria), Health Information Technology for Economic and Clinical Health (HITECH) Act 2009 (United States), Health Services Organisation Act 2001 (Estonia), Medicare Access and CHIP Reauthorization Act 2015 (United States), NHS Community Care Act 2010 (United Kingdom | 18 | 41 |
| Data Protection Laws (Supranational) | Data Protection Directive (European Union), General Data Protection Regulation (European Union) | 13 | 31 |
| National Health Service Laws | Care Act 2014 (United Kingdom), Health Act 2004 (Ireland), Health and Social Care Act 2012 (United Kingdom), Hopital, Patient, Sante et Territores Act 2009 (France), National Health Service Act 2006 (United Kingdom), Public Health Act 2004 (France), Public Health Act 2016 (France) | 10 | 20 |
| Research Ethics Laws | Act on Health Ethics 2018 (Denmark), Belmont Report (United States), Caldicott Principles (United Kingdom), Common Rule 1991 (United States), Danish Code of Conduct for Research Integrity (Denmark), Declaration of Helsinki (International), Human Genes Research Act 2000 (Estonia), NHMRC National Statement on Ethical Conduct in Human Research (Australia), Law on Medical Research 488/1999 (Finland), Law on Medical Research 794/2010 (Finland). | 10 | 20 |
| National Electronic Health Record Guidelines (Supranational) | Directive 2011/24/EU (European Union), eHealth Action Plan (European Union), Regulation 859/2003/EC (European Union), Regulation 1231/2010/EC (European Union) | 6 | 18 |
| Healthcare Ethics Law | Health Care Law 1326/2010 (Finland), Law on Practising Physicians 1998 (China), Lisbon Declaration on the Rights of the Patient 1981 (International), Medical Care Act (Taiwan), Medical Registration Act 2014 (Singapore), Medical Treatment Contract Act 1994 (Netherlands), NHS Code of Practice on Confidential Information (United Kingdom), Physicians Act (Taiwan), Regulations on the Urgent Handling of Public Health Emergencies (China), Medical Council Guidelines (Singapore) | 6 | 12 |
| Patient Rights Law | Act 18355 from August 15, 2008 (Uruguay), Act of Reading health or social care client information 159/2007 (Finland), Act on the status and rights of patients 785/1992 (Finland), Law of 22 August 2002 on Patient Rights (Belgium), Law of Client Rights Regarding Electronic Data Processing in Healthcare (Netherlands) | 4 | 9 |
| Pharmaceutical and Device Law | Directive 93/42/EEC (European Union), Directive 2012/52/EU (European Union), Food and Drug Administration Amendments Act 2007 (United States), Medicare Prescription Drug Improvement and Modernization Act 2003 (United States) | 6 | 7 |
| Criminal Law | Computer Misuse Act 1990 (United Kingdom), Criminal Code 39/1889, 940/2008 (Finland), Criminal Procedure Law 2012 (China) | 3 | 6 |
| Torts Law | Tort Law 2009 (China), Regulations on the Handling of Medical Malpractice (2002) | 1 | 6 |
| Electronic Signatures Law | Act 18600 from September 21, 2009 (Uruguay), Directive 1999/93/EC (European Union), Legislative Decree 82/2005 (Italy) | 3 | 4 |
| Archives Law | Archives Act 2003 (Malaysia) | 1 | 3 |
| Intellectual Property Law | Copyright Act 1987 (Malaysia) | 1 | 1 |

## Table 3. Stakeholders

| Name | Terms | Docs (n) | Segments (s) |
|---|---|---|---|
| Patient and Lay Public | Public; patient; user; citizen | 85 | 2006 |
| Primary Healthcare Providers | General Practitioner; physician; doctor; nurse | 80 | 2051 |
| Healthcare Organisations | healthcare organisations; hospitals | 78 | 924 |
| Government | Government; politician | 71 | 624 |
| Academics | Academics; researchers; scientist's; institutes; analyst; schools; university; professor | 71 | 548 |
| Information Communications Technology | IT professionals; IT providers; IT administrators; technicians; software developers; software vendors; data controllers; application developer; developer | 37 | 184 |
| Legal | Lawyer; lawmaker; attorney; legal office; legal representative; legal executive; legal practitioner; law reform commission; legal professionals | 37 | 92 |
| Pharmaceutical | Pharma; pharmacists; pharmaceutical companies; drug companies; novartis; roche | 33 | 101 |
| Data Protection | Data protection authority; data protection group; data protection working party; data protection agency; data protection; data protection expert; data access committees (DAC); privacy advocates; research ethics committees; research ethics expert | 27 | 122 |
| Insurance Organisation | Insurance companies; insurance organisations | 27 | 65 |
| Specialists | Specialists; radiologists; specialist healthcare providers | 23 | 48 |
| Administrative | Administrator; administrative staff; health administrator; healthcare managers; veterans health administrator; biobank administrators | 18 | 59 |
| Media | Media; newspaper; | 18 | 84 |
| Allied Health | Allied health professionals; occupational groups; occupational therapists; physiotherapists; psychiatrists; psychologists | 13 | 27 |
| Private Healthcare | Commercial; corporations; private companies; private Sector; private sector organisations; private sector organizations; private sector stakeholders | 10 | 75 |
| Non-profit | Non-government organisation; non-government organization; Not-for-profit organisation; Not-for-profit organization | 10 | 20 |
| Social Work | Social security; social workers | 5 | 6 |

## Table 4. Benefits of eHealth systems

| Benefit | Docs (n) | Segments (s) |
|---|---|---|
| Improving healthcare processes & delivery | 57 | 156 |
| Improving access and exchange of information | 57 | 144 |
| Improving quality of care | 48 | 107 |
| Supporting and informing research and health policy | 43 | 92 |
| Patient empowerment and engagement | 40 | 91 |
| Improving patient safety and security | 45 | 75 |
| Reducing costs | 27 | 51 |
| Enabling better monitoring and evaluation of services | 13 | 16 |
| Generally increases ability to address challenges | 5 | 10 |

## Table 6

| **Challenges** | | **Docs (n)** | **Segments (s)** |
|---|---|---|---|
| Implementation | Conflicting Stakeholder Interests, Difficulty Demonstrating Benefits, Financial Issues, Government Policy and Political Issues, Implementation Problems | 57 | 156 |
| Ethical-Legal | Privacy Concerns, Research Ethics Concerns, Patient Autonomy, Medical Liability, Other Legal Challenges | 57 | 144 |
| Data Related | Data Availability, Information Quality, Lack of Interoperability | 48 | 107 |
| Engagement | Access to Benefit and Equity, Communication Issues, Lack of Patient Uptake, Lack of Physician Uptake | 43 | 92 |
| Technical | Low Usability, Security Concerns, Technical Flaws | 40 | 91 |

## Table 8

| **Contributors to Success** | | **Docs (n)** | **Segments (s)** |
|---|---|---|---|
| Socio-cultural | - Promote cultural and normative acceptance at all levels of the system and amongst wider society<br>- Promote multidisciplinary collaborations and cooperation<br>- Foster support, trust and ownership of the system amongst all actors<br>- Incentivise engagement & provide training / guidance<br>- Sharing experiences / lessons learned | 65 | 307 |
| System-based | - Create a system that is flexible, intuitive, interoperable, transparent, accessible, de-centralised, safe and reliable<br>- Robust data governance structures and context-sensitive security<br>- Ensure data accuracy<br>- Ensure consent mechanisms promote informational autonomy | 64 | 212 |
| Legal | - Abide by standards, protocols, guides & recommendations | 45 | 134 |
| Implementation | - Good project management, clearly defined goals, realistic timelines<br>- Perform evaluations early and iteratively<br>- Pay attention to conflicts of interest and disruptions to workflows | 40 | 149 |
| Resources | - Ensure availability of required resources such as funding, infrastructure, technologies, and qualified human resources | 32 | 82 |